\documentclass[aps,prd,nofootinbib,twocolumn,superscriptaddress,preprintnumbers]{revtex4-1}

\usepackage{mathrsfs,graphicx,rotating,amsmath,amsfonts,mathtools,booktabs,amssymb,wasysym,dsfont,hyperref,slashed}

\newcommand{\Ra}{R_{\rm{AdS}}}
\DeclarePairedDelimiter\abs{\lvert}{\rvert}%
\newenvironment{aleq}
    {\begin{equation}\begin{aligned}}
    {\end{aligned}\end{equation}\ignorespacesafterend}

\usepackage{soul,xcolor}
\begin{document}

\title{Integer Conformal Dimensions for Type IIA Flux Vacua}
\author{Fien Apers}
\affiliation{Rudolf Peierls Centre for Theoretical Physics\\ Beecroft Building, Clarendon Laboratory, Parks Road, University of Oxford, OX1 3PU, UK}
\author{Joseph P. Conlon}
\affiliation{Rudolf Peierls Centre for Theoretical Physics\\ Beecroft Building, Clarendon Laboratory, Parks Road, University of Oxford, OX1 3PU, UK}
\author{Sirui Ning}
\affiliation{Rudolf Peierls Centre for Theoretical Physics\\ Beecroft Building, Clarendon Laboratory, Parks Road, University of Oxford, OX1 3PU, UK}
\author{Filippo Revello}
\affiliation{Rudolf Peierls Centre for Theoretical Physics\\ Beecroft Building, Clarendon Laboratory, Parks Road, University of Oxford, OX1 3PU, UK}
    
\begin{abstract}
We give a concise argument that supersymmetric AdS type IIA DGKT flux vacua on general Calabi-Yaus have, interpreted holographically, integer conformal dimensions for low-lying scalar primaries in the dual CFT. 
These integers are independent of any compactification details, such as the background fluxes or triple intersection numbers of the compact manifold. For the K\"ahler moduli and dilaton, there is one operator with $\Delta = 10$ and $h^{1,1}_{-}$ operators with $\Delta =6$, whereas the corresponding axions have $\Delta = 11$ and $\Delta = 5$.
For the complex structure moduli, the $h^{2,1}$ saxions have $\Delta =2$, and the axions $\Delta =3$. We give a tentative discussion of the origin of these integers and effects that would modify these results.
\end{abstract}

\maketitle

\noindent \textbf{Introduction}\\

Moduli stabilisation is one of the necessary pre-requisites for compactified string theories to describe a 4-dimensional world with a hierarchical separation of scales between this and extra-dimensional physics.
The traditional tools to construct such vacua, effective field theory and dimensional reduction, 
have come under close scrutiny with the advent of the Swampland program \cite{Vafa:2005ui,Ooguri:2006in,Palti:2019pca}. For this reason, it was proposed in \cite{Conlon:2018vov,Conlon:2020wmc,Conlon:2021cjk,Apers:2022zjx} that holography might provide an independent approach to test the consistency of AdS vacua, by analysing the properties of the putative dual CFTs. 

One of the most studied scenarios of moduli stabilisation featuring scale separation are the type IIA DGKT flux vacua \cite{DeWolfe:2005uu}, and scale separated IIA vacua have received considerable attention lately \cite{Font:2019uva,Junghans:2020acz,Marchesano:2020qvg,Marchesano:2020uqz,Cribiori:2021djm,Apers:2022zjx,Farakos:2020phe}, as the scale separation property has been conjectured to lie in the Swampland \cite{Gautason:2018gln,Lust:2019zwm,Buratti:2020kda}.\footnote{However, both the AdS moduli conjecture of \cite{Gautason:2018gln} and a refined version of the Strong AdS distance conjecture \cite{Lust:2019zwm} presented in \cite{Buratti:2020kda} are consistent with DKGT.} From a holographic point of view, scale separated vacua also admit a clear interpretation in terms of CFTs with a gapped spectrum, which could also point towards inconsistencies \cite{Collins:2022nux}. Therefore, DGKT provides a particularly interesting example to study in this context (see \cite{Aharony:2008wz} for an early investigation).

In this Letter we give a concise derivation of 
the mass matrix for general IIA DGKT flux vacua and show that, interpreted holographically, it has an extremely simple form. In particular, the conformal dimensions of scalar operators dual to the moduli are both integers and also highly degenerate. These results are suggestive of a hidden structure that is best understood from a dual CFT perspective.

Our argument establishes in full generality results hinted at in \cite{Conlon:2021cjk} and \cite{Apers:2022zjx}, where these results was found for simple specific examples of DGKT. Although the argument here is more concise, a derivation of the mass matrix for general DGKT vacua (although without a link to conformal dimensions) also appears spread across the two papers \cite{Herraez:2018vae, Marchesano:2019hfb}. \\

\noindent \textbf{DGKT flux vacua}\\

In a type IIA setting, moduli stabilisation with fluxes can be achieved at tree-level due to the simultaneous presence of the NS-NS $B_2$ form and the R-R odd $p$-forms. In particular, this is obtained by compactifying massive IIA string theory on Calabi-Yau orientifolds with O6-planes (for which the effective field theory was derived in \cite{Kachru:2004jr,Grimm:2004ua}), which allows for stable AdS$_4$ vacua in the controlled limit of large volumes and weak string coupling \cite{DeWolfe:2005uu}. This was demonstrated in \cite{DeWolfe:2005uu} for a fully explicit example based on a $T^6/ \mathds{Z}_3 \times \mathds{Z}_3$ orientifold with no complex structure (CS) moduli ($h^{2,1}=0$). There, the axio-dilaton, K\"ahler moduli and axions are all stabilised by fluxes. 

Similar results apply for a generic $\mathcal{N}=1$ orientifold, where it is again possible to fix all moduli (except for the flat directions corresponding to the complex structure axions). Following the orientifold projection, one is left with $h^{1,1}_-$ K\"ahler moduli and $h^{2,1}$ CS moduli, which can be described with the formalism of $\mathcal{N}=1$ supergravity. The K\"ahler moduli can be expressed in terms of the complex scalar fields
\begin{equation}
t_a = b_a+ i v_a \quad \quad a = 1, \ldots \, h_{-}^{1,1}
\end{equation}
where the $v_a$ are volumes of the 2-cycles ($\int J$) and the $b_a$ axions arise from dimensional reduction of the $B_2$ form on this cycle. Their K\"ahler potential is given by
\begin{equation}\label{kahler}
    K^K = - \log \left( \dfrac{4}{3} \mathcal{V}\right),
\end{equation}
where the volume is\footnote{Our definition differs from the proper volume by $\text{Vol}=\mathcal{V}/6$. }
$\mathcal{V} = \kappa_{abc} v_a v_b v_c.$ 
This K\"ahler potential satisfies the well known no-scale relations
\begin{equation}\label{noscale1}
   K^{ab} K_a K_{b}=3 \quad \quad K^{ab} K_b = -v_{a}.
\end{equation}
The complex structure moduli, together with the axio-dilaton, can be packaged into $h^{2,1}+1$ complex fields
\begin{equation}\label{eq:csm}
\begin{aligned}
&N_{k} = \frac{\xi_k}{2}+  i \text{Re}(C Z_k) \quad \quad k = 0 \ldots \tilde{h} \\ &T_{\lambda} = i \tilde{\xi_\lambda} -2 \text{Re}(C g_\lambda) \quad \quad  \lambda = \tilde{h} + 1  \ldots h^{2,1}
  \\
\end{aligned}
\end{equation}
where the $Z_k,g_{\lambda}$ and the $\xi_k, \tilde{\xi}_{\lambda}$ are the coefficients of the holomorphic 3-form $\Omega$ and of $C_3$ 
\begin{equation}
\begin{aligned}
& \Omega=Z_{\hat{K}} \alpha_{\hat{K}}-g_{\hat{L}} \beta_{\hat{L}}\\
& C_{3}=\xi_{\hat{K}} \alpha_{\hat{K}}-\tilde{\xi}_{\hat{L}} \beta_{\hat{L}},
\end{aligned}
\end{equation}
expanded in a symplectic basis $\{\alpha_{\hat{K}},\beta_{\hat{L}}\}$ of $H^{3}$. The basis can be split into an odd part $\{\alpha_{\lambda},\beta_{k} \}$ and even part $\{\alpha_{k},\beta_{\lambda} \}$; it is only the components with respect to the latter which survive the orientifold projection and appear in (\ref{eq:csm}). We have also introduced the compensator $C \equiv e^{-D+K_{\text{cs}}/2}$, where the 4d dilaton $D$ is related to the 10d one by $e^D = e^{\phi}/\sqrt{\text{Vol}}$ and
\begin{equation}
 K_{\text{cs}} = -  \log (i \int \Omega \wedge \bar{\Omega}). 
\end{equation}
Their K\"ahler potential is given by
\begin{equation}\label{csk}
    K^Q = - 2 \log \left( 2 \int \text{Re}(C \Omega) \wedge * \text{Re}(C \Omega) \right) = 4 D
\end{equation}
and also satisfies a no-scale relation (see appendix C of \cite{Grimm:2004ua}), namely
\begin{equation}\label{noscale2}
   K^{N_k \bar{N}_k} K_{N_k} K_{\bar{N}_k}+ K^{T_{\lambda} \bar{T}_{\lambda}} K_{T_{\lambda}} K_{\bar{T}_{\lambda}} = 4.
\end{equation}

In a generic setting, both the NS-NS 3-form field strength $H_3$ and the R-R field strengths $F_0, F_2, F_4, F_6$ can thread fluxes through the internal manifold. Following the notation of \cite{Grimm:2004ua}, the background fluxes can be expressed in a basis of the appropriate cohomologies as
\begin{equation}
\begin{aligned}
H_{3}=q_{\lambda} \alpha_{\lambda}-p_{k} \beta_{k},& \quad F_{2}=-m_{a} w_{a}, \quad F_{4}=e_{a} \tilde{w}^{a},\\
\quad &F_{0}=m_{0}, \quad F_6 = e_0. \\
\end{aligned}
\end{equation}
The even 2-forms $\{ w_{a} \}$ span a basis of $H^{1,1}_+$, while their duals $\{ \tilde{w}^{a} \}$ are a basis of $H^{2,2}_+$. We briefly remark that the presence of the $F_2$ and $F_6$ fluxes is not needed to achieve moduli stabilisation, but we nevertheless include them for full generality.\footnote{In fact, it usually leads to solutions which are either equivalent or qualitatively similar to the ones without $F_2$ or $F_6$ flux.}

The resulting superpotential is given by
\begin{equation}\label{superpotential}
\begin{aligned}
    W = e_0+ e_at^a &+ \dfrac{1}{2}\kappa_{abc}m_a t_b t_c - \dfrac{m_0}{6} \kappa_{abc} t_a t_b t_c\\ &-2 p_{k} N_{k}-i q_{\lambda} T_{\lambda}.
    \end{aligned}
\end{equation}

A crucial simplification is that the superpotential (\ref{superpotential}) depends only on a linear combination of the complex structure moduli which, after a (holomorphic) rotation in field space, can be effectively taken to be a single modulus. When $h^{2,1}=0$ this direction reduces entirely to the axio-dilaton,\footnote{In that case, the imaginary part of $S$ will be related to the 4d dilaton by $s=e^{-D}/\sqrt{2}$.} and hence it will be denoted as
\begin{equation}
S \equiv \xi+ i s, \quad \quad \text{with} \quad \quad W \supset -2 p S
\end{equation}
as the form for the superpotential. Combined with the fact that the K\"ahler potential factorises as a sum of two independent terms, this ensures a decoupling between the two sectors. Other than $S$, there will now be $h^{2,1}$ complex structure moduli 
\begin{equation}
U_{\alpha} \equiv a_{\alpha}+ i u_{\alpha}, \quad \quad \alpha = 1, \ldots h^{2,1}
\end{equation}
which do not appear in the superpotential $W(t_a,S)$.

Assuming the tadpole conditions are satisfied (as they must be), the scalar potential then takes the standard $\mathcal{N}=1$ supergravity form \cite{Grimm:2004ua,DeWolfe:2005uu}
\begin{equation}\label{eq:potential}
V =  e^K \Bigg( \sum_{t_i,U_{\alpha},S} K^{i \bar{j}} D_i W D_{\bar{j}} \overline{W} - 3 \abs{W}^2 \Bigg),
\end{equation}
Supersymmetric vacua occur with vanishing F-terms:
\begin{equation}\label{fterm}
D_{t_i} W = 0 \quad \quad D_S W = 0 \quad \quad D_{U_{\alpha}} W = 0,
\end{equation} 
which also ensures extrema of the potential (\ref{eq:potential}).
For the superpotential \eqref{superpotential} and K\"ahler potential \eqref{kahler}, these imply the following relationships for the K\"ahler moduli at the minimum \cite{DeWolfe:2005uu}:
\begin{equation}\label{eq:vs}
\begin{aligned}
 3 & m_0^2 \kappa_{abc} v^b v^c +10 m_0 e_a +5\kappa_{abc}m^b m^c = 0, \\
 & s = -\frac{2\, m_0}{15\, p } \mathcal{V}, \quad \, b_a = \frac{m_a}{m_0}, \quad \, \xi = -\frac{e_0}{2 p}. \\
 \end{aligned}
\end{equation}
This implies that at the minimum \cite{DeWolfe:2005uu}
\begin{equation}\label{Wmin}
    W = \dfrac{2i}{15} m_0 \mathcal{V}.
\end{equation}
Eqs (\ref{eq:vs}) and (\ref{Wmin}) imply the second derivatives of the superpotential take the simpler form
\begin{aleq}\label{Wders}
    \partial_{v_a} \partial_{v_b} W & = - \partial_{t_a} \partial_{t_b} W \\
    &= -\kappa_{cab}m_c +m_0 \kappa_{abc} t_c 
     = i m_0 \kappa_{abc} v_c \\
   & = i
 \dfrac{m_0 \mathcal{V}}{6} (K_a K_b - K_{ab}) = \dfrac{5}{4} (K_a K_b - K_{ab})W.\\
\end{aleq}
when Eqs (\ref{eq:vs}) are satisfied. On the other hand,
the F-term condition for the complex structure moduli simply implies
\begin{equation}\label{eq:csf}
   K_{U_{\alpha}} = \frac{1}{2 i} K_{u_{\alpha}} = 0,
\end{equation}
ensuring the absence of mixing with the K\"ahler moduli and $S$ for fluctuations around the minimum. In addition, (\ref{eq:csf}) turns the no-scale relation (\ref{noscale2}) into 
\begin{equation}
K^{ss} K_s K_s = 4.
\end{equation}

\noindent \textbf{Mass matrices for the K\"ahler moduli and dilaton}\\

To obtain the mass matrix, we compute the second derivatives of the potential \eqref{eq:potential},
\begin{aleq}\label{secondder}
&\partial_m \partial_n V = -3e^K(K_{mn} \abs{W}^2 + K_m \partial_n\abs{W}^2 +\partial_m \partial_n\abs{W}^2) \\
&+ 2e^K K^{i \Bar{j}} (\partial_n D_i W)(\partial_m D_{\Bar{j}} \overline{W}),\\
\end{aleq}
in which terms proportional to a first derivative of the potential or $D_i W$ are dropped, and $m,n$ denote saxions.\\
To evaluate the derivatives in the first line of \eqref{secondder}, we note that
the conditions \eqref{fterm} imply that
\begin{equation}\label{Wa}
    \partial_m \abs{W}^2 = -K_m \abs{W}^2,
\end{equation}
and the second derivatives of $\abs{W}^2$ equal
\begin{aleq}\label{derivatives}
&  \partial_s^2 \abs{W}^2= \dfrac{1}{2}K_s^2 \abs{W}^2, \\
& \partial_s \partial_b \abs{W}^2 = \dfrac{1}{2} K_s K_b \abs{W}^2, \\
&  \partial_{b}\partial_{a}\abs{W}^2 = 2 (\partial_{v_a} \partial_{v_b} W ) \overline{W} + 2 (\partial_{v_a} W) ( \partial_{v_b} \overline{W})\\
&= \left( 3K_a K_b - \dfrac{5}{2} K_{ab} \right) \abs{W}^2.
\end{aleq}
For the second line in \eqref{secondder}, we have
\footnote{Note that because
$K = K(t_a + \overline{t}_a, S+ \overline{S}),$
the derivatives w.r.t. the complex moduli are
$
\partial_{ t_a} K= \partial_{ v_a}K/2i,\   \partial_{ S}K = \partial K_{s}/2i.
$
The superpotential is holomorphic and so
$
\partial W_{t_a} = -i\partial_{v_a}W,\  \partial_S W= -i \partial_s W.
$}
\begin{aleq}
&K^{i \bar{j}}(\partial_n D_iW)(\partial_m D_{\bar{j}} \overline{W})\\ 
&= K^{i \bar{j}} (W_{in} + K_{in}W-\dfrac{K_i K_n}{2} W)\\
&\times(\overline{W}_{\overline{j}m}+ K_{\Bar{j}m}\overline{W} -\dfrac{K_{\Bar{j}}K_m}{2} \overline{W})\\
&= K^{i \Bar{j}} W_{in} \overline{W}_{\Bar{j}m} + 2 K^{i \Bar{j}} W_{in} K_{\Bar{j}m}\overline{W}-  K^{i \Bar{j}} W_{in} K_{\Bar{j}} K_m \overline{W}\\
&+\Big[ K^{i \Bar{j}} K_{in} K_{\Bar{j}m}  -  K^{i \Bar{j}} K_{in} K_{\Bar{j}} K_m \\ &+ \dfrac{1}{4} (K^{i \Bar{j}} K_{i}K_{\Bar{j}}) K_n K_m\Big] \abs{W}^2\\
&=  K^{i \Bar{j}} W_{in} \overline{W}_{\Bar{j}m} + 4W_{nm} \overline{W} -  K^{i \Bar{j}} W_{in} K_{\Bar{j}} K_m \overline{W}\\
&+ \left[K_{nm} + \dfrac{3}{4}K_n K_m\right] \abs{W}^2.
\end{aleq}
If $n,m=a,b$ are both size moduli,
\begin{aleq}
 K^{i \Bar{j}} W_{in} \overline{W}_{\Bar{j}m} = \dfrac{25}{16}(K_a K_b + K_{ab}) \abs{W}^2,
\end{aleq}
and if $n=a$ is a size modulus
\begin{aleq}
 K^{i \Bar{j}} W_{in} K_{\Bar{j}} K_m \overline{W} = 5 K_a K_b \abs{W}^2
\end{aleq}
by substituting \eqref{Wders}, and otherwise these terms vanish.
Combining these leads to
\begin{equation}
   \partial_{a} \partial_{b} V = e^K ( 9 K_{ab} + 8 K_a K_b) \abs{W}^2,
\end{equation}
\begin{equation}\label{sa}
   \partial_{a} \partial_s V = e^K (-2 K_a K_s) \abs{W}^2,
\end{equation}
\begin{equation}
    \partial_s^2 V = e^K (-K_{ss} + 3 K_s^2) \abs{W}^2.
\end{equation}
Hence, the mass matrix for the moduli in AdS units is given by
\begin{equation}
    \Ra^2 \partial_m \partial_n V = \begin{pmatrix}
      9 K_{ab} + 8 K_a K_b & -2 K_a K_s\\
      -2 K_a K_s & -K_{ss} + 3 K_s^2
    \end{pmatrix},
\end{equation}
where $\Ra^2 = -3/V_{min}$.
We can proceed similarly with expression \eqref{secondder} for the axions $b_a,\xi$, where now all derivatives of the K\"ahler potential will vanish, and for the superpotential derivatives we have
\begin{aleq}
\partial_{b_a} W = -i \partial_{v_a} W, \quad \partial_{\xi} W = -i \partial_s W.
\end{aleq}
This results in
\begin{equation}
    \partial_{b_a} \partial_{b_b} V = e^K(5K_{ab} + 12 K_a K_b) \abs{W}^2,
\end{equation}
\begin{equation}
    \partial_{b_a} \partial_{\xi} V = -3 e^K K_a K_s \abs{W}^2,
\end{equation}
\begin{equation}
    \partial_{\xi}^2 V = 2 e^K K_s^2 \abs{W}^2,
\end{equation}
and so the mass matrix is 
\begin{equation}
    \Ra^2 \partial_m \partial_n V = \begin{pmatrix}
      5 K_{ab} + 12 K_a K_b & -3 K_a K_s\\
      -3 K_a K_s & 2 K_s^2
    \end{pmatrix}.   
\end{equation}
These expressions match with the results obtained in \cite{Marchesano:2019hfb} (see in particular Appendix B). However, this computation provides a more compact way to arrive at these results, using the ordinary $\mathcal{N}=1$ structure of the potential and without having to rely on the formalism developed in \cite{Herraez:2018vae,Marchesano:2019hfb,Marchesano:2020uqz}. Moreover, the derivation provided above makes it clear that the only properties feeding into the final result are the no-scale relations for the K\"ahler potential and the specific form of the superpotential leading to (\ref{Wmin}) and (\ref{Wders}). \\

\noindent \textbf{Complex structure moduli}\\

Let us briefly review the arguments that lead both to integer conformal dimensions for the complex structure moduli (which have already been discussed (implicitly or explicitly) in \cite{Conlon:2006tq,Conlon:2021cjk}) and the absence of any mixing with the K\"ahler sector. To establish the latter, one can follow the same steps as the ones leading to \eqref{sa}, finding that
\begin{equation}
    \partial_{u_\alpha} \partial_{v_b} V = -2K_{u_\alpha} K_{v_b} e^K \abs{W}^2 = 0
\end{equation}
because of (\ref{eq:csf}). Similarly,
\begin{aleq}
 \partial_{u_\alpha} \partial_{s} V &= -3e^K(-K_{u_\alpha} K_s  + \dfrac{1}{2} K_{u_\alpha} K_s) \abs{W}^2\\
 & + \dfrac{3}{2} K_{u_\alpha}K_s e^K \abs{W}^2 = 0.
\end{aleq}
Within the CS sector, non-diagonal terms are also absent
\begin{aleq}
 \partial_{u_\alpha} \partial_{u_\beta} V &= 3e^KK_{u_\alpha} K_{u_\beta} \abs{W}^2 = 0.
\end{aleq}
It was shown in \cite{Conlon:2006tq} that the only non-vanishing contribution then comes from
\begin{equation}
    \partial_{u_{\alpha}}\partial_{u_{\alpha}} V = - K_{u_{\alpha}u_{\alpha}} e^K \abs{W}^2,
\end{equation}
fixing all masses to the universal value of
\begin{equation}\label{eq:css}
m^2_u  = -2/3 V_{\text{min}}= -2/\Ra^2.
\end{equation}
As the axions appear in neither the K\"ahler potential nor the superpotential, they are all massless.\\

\noindent \textbf{Spectrum}\\

Let us define the two block matrices
\begin{equation}
K_{mn} =
\begin{pmatrix}
 K_{ab} & 0 \\ 
0 & K_{ss} \\ 
\end{pmatrix}
\quad
L_{mn} =
\begin{pmatrix}
 4 K_{a} K_{b} & -K_a K_s \\ 
-K_a K_s & \frac{K_s^2}{4} \\
\end{pmatrix},
\end{equation}
where $K$ is simply the K\"ahler metric for the moduli (in real components) evaluated at the minimum of the potential. Then, from the results of the previous section the Hessians of the moduli and axion potential can be expressed as
\begin{equation}
H^M_{mn} \equiv \Ra^2 \partial_m \partial_n V = 9 K_{mn}+2L_{mn}
\end{equation}
and
\begin{equation}
H^A_{mn} \equiv \Ra^2 \partial_m \partial_n V =
5 K_{mn}+3L_{mn}.
\end{equation}
respectively. The physical masses can be obtained as the eigenvalues of $ M = 2K^{-1} H$, which is related to the mass matrix
\begin{equation}
\tilde{M}= 2 \sqrt{K^{-1}}^T H \sqrt{K^{-1}}
\end{equation}  
by a similarity transformation. 

In principle, the stabilized values of the $v_i$'s and the dilaton can be extracted by the system of $h^{1,1}_{-} +1$ equations in the first line of (\ref{eq:vs}). Knowing the value of the axions at the minimum, it would then be possible to write the mass matrix as a function of the flux data only. Although the equations cannot be solved explicitly for arbitrary triple intersection numbers $\kappa_{abc}$ and fluxes $e_a,m_a$, the ‘‘implicit" form of the matrix derived above is not only sufficient for our purposes, but it is more helpful to elucidate the surprising simplifications in the final expressions.

To compute the eigenvalues of $K^{-1} L$, one can begin by noticing that any vector of the form
\begin{equation}
x = (x_i,x_0) \quad \text{with} \quad x_0\, K_s = 4 K_a x^a
\end{equation}
satisfies $K^{-1}\,L\, x = 0$. This essentially descends from the fact that $M$ can be expressed as a tensor product $L_{mn} = A_m A_n$, with $A_m= (2K_a,-K_s/2)$. One therefore has a basis of $h^{1,1}_{-}$ eigenvectors with an eigenvalue of $\lambda_x^i = 0$. Furthermore, the no-scale relations (\ref{noscale1})-(\ref{noscale2}) imply that the vector
\begin{equation}
y= (v_i,K_s),
\end{equation}
is also an eigenvector of $L$, with eigenvalue $\lambda_y = 13$.

Then, there is a basis in which the mass matrices in AdS units read
\begin{equation}
M^M_{mn} = 18 \delta_{mn} + 52 \,\delta_{1,m}
\end{equation}
\begin{equation}
M^A_{mn} = 10 \delta_{mn} + 78 \,\delta_{1,m},
\end{equation}
also in agreement with the results of \cite{Marchesano:2019hfb}.
Through the relation $\Delta(\Delta-d) = m^2 \Ra^2$, they correspond to conformal dimensions of 
\begin{equation}\label{eq:delta1}
\Delta_1 = 10, \quad \quad \Delta_{2...h_-^{1,1}+1} = 6
\end{equation}
for the saxions, and
\begin{equation}\label{eq:delta2}
 \Delta_1 = 11, \quad \quad \Delta_{2...h_-^{1,1}+1} = 5.
\end{equation}
for the corresponding axions. 
Similar conclusions can be drawn for all the complex structure moduli as well. From Eq. (\ref{eq:css}), there are in principle two distinct possibilities for the saxions as both solutions to $\Delta(\Delta-d) = m^2 \Ra^2$ satisfy the unitarity bound. However, $\mathcal{N}=1$ superconformal symmetry excludes the option $\Delta_{u_{\alpha}} = 1$,\footnote{It would be interesting to understand how supersymmetry implements this from an AdS perspective, where the choices correspond to different boundary conditions for the scalar field. } as the axion and saxion are in the same $3d$ $\mathcal{N}=1$ supermultiplet \cite{Cordova}. Combined with the fact the complex structure axions are all massless, one arrives at the conclusion that
\begin{equation}\label{eq:delta3}
     \Delta_{u_{\alpha}} = 2, \quad \Delta_{a_{\alpha}} =3.
\end{equation}
for all complex structure moduli $U_{\alpha}$, where $\alpha = 1...h^{2,1}$. 
\\

\noindent \textbf{Conclusions}\\

We have shown how - if we assume they actually exist - the spectrum of the CFT$_3$ duals to supersymmetric DGKT type IIA flux vacua is characterized by a spectrum of integer conformal dimensions for the low-lying primaries. This results generalises an observation first made in \cite{Conlon:2021cjk} for the simplest case of a $T^6/ \mathds{Z}_3 \times \mathds{Z}_3$ orientifold, and later extended in \cite{Apers:2022zjx} to a wide range of examples. 

For a generic modulus, the conformal dimension is given by one of the very few integer values in Eqs (\ref{eq:delta1})-(\ref{eq:delta3}). It is surprising to see how these numbers do not bear any trace of the microscopic details of the compactification, i.e. the values of the fluxes and the geometry of the orientifold. Such a simple and universal behaviour is suggestive of the fact that the holographic perspective offers both a novel and insightful viewpoint on these constructions.

Despite the simplicity of the result, the calculation itself does not seem to yield any clear insight to why such a striking property should hold. Its outcome relies on the no-scale relations for the K\"ahler potential, as well as some more specific conditions involving the superpotential and its derivatives, Eqs (\ref{Wmin}) and (\ref{Wders}). 

One might wonder whether supersymmetry plays a role - already at the $\mathcal{N}=1$ level, the structure of the superconformal multiplets implies that the conformal dimensions of axions and saxions should differ by one. A further observation is that no-scale relations partially depend on the factorisation of the K\"ahler potential into a sum of two terms (depending only on the K\"ahler and complex structure moduli respectively), which could be seen as a remnant of a more constraining $\mathcal{N}=2$ supersymmetry of the original supergravity theory, where the K\"ahler potential factorises into hypermultiplet and vector multiplet terms. 

Generally, one would expect that $\mathcal{N}=1$ corrections would break this factorisation and introduce mixing between the K\"ahler and complex structure moduli. Examples of such corrections would be $\alpha'$ and $g_s$ corrections which, generically, would not preserve the $\mathcal{N}=2$ structure. The DGKT vacua have a scale-separated limit of large volume and weak coupling - one would expect this limit to suppress such corrections, with an asymptotic limit in which corrections to the integer conformal dimensions vanish asymptotically.

More generally, it would be extremely fascinating to come up with a deeper explanation of either the integer dimensions or the large degeneracies amongst the operators. A related question one might hope to address in the future is whether any similar structure holds for the non-supersymmetric vacua as well, as suggested by the examples in \cite{Conlon:2021cjk,Apers:2022zjx}.

 In a long term perspective, an interesting direction would involve trying to understand if one could either prove or disprove the consistency of these (or similar) 4d vacua using ideas from AdS/CFT.
\\

\noindent \textbf{Acknowledgements}\\

We would like to thank Miguel Montero, Thomas Van Riet and Timm Wrase for conversations related to their recent work \cite{Apers:2022zjx}, and Pietro Ferrero for discussions. We thank STFC for funding via the Consolidated Grant. SN acknowledges funding support from the China Scholarship Council-FaZheng Group- University of Oxford. FR is supported by the Dalitz Graduate Scholarship, jointly established by the Oxford University Department of Physics and Wadham College.
FA acknowledges the Clarendon Fund Scholarship in partnership with the Scatcherd European Scholarship, Saven European Scholarship and the Hertford College Peter Howard Scholarship.

\end{document}